\documentstyle[]{article}
\title{ Vortices in superconducting films - statistics
        and fractional quantum Hall effect }
\author{Jacek Dziarmaga  \\
        Jagellonian University, Institute of Physics, \\
        Reymonta 4, 30-059 Krak\'ow, Poland
        \thanks{address after October 1, 1995 -
                Department of Mathematical Sciences, University of Durham,
                South Road, Durham, DH1 3LE, UK}
        \thanks{e-mail address: J.P.Dziarmaga@durham.ac.uk}}
\date{November 6, 1995}
\begin{document}
\maketitle
   \begin{abstract}
   We present a new derivation of the Berry phase picked up during exchange
of parallel vortices. This derivation is based on the
Bogolubov - de Gennes formalism. The origin of the Magnus force is also
critically reanalised. The Magnus force can be interpreted as interaction
with effective magnetic field. The efective magnetic field may be even of
the order $10^{6} T/\AA$. We discuss a possibility of the FQHE in vortex
systems. As the real magnetic field is varied to drive changes
in vortex density, the vortex density will prefer to stay at some quantised
values. The mere existence of the FQHE does not depend on vortex
quantumstatistics although the pattern of the plateaux does. We also
discuss how the density of anyonic vortices can lower the effective strengh
of the Magnus force, what might be observable in measurements of Hall
resitivity.
   \end{abstract}

\subsection*{To appear in Physical Review B.}
\subsection*{cond-mat/9508140}
\hspace*{0.5cm}

\section{ Introduction }

   In our recent paper \cite{anyon} we have pointed out the possibility
that vortices in superconducting films might be anyons. There is a
classic paper by Haldane and Wu \cite{hawu}, which demonstrates that
vortices in superfluid helium layers are not anyons because there
is no well defined dilute limit.
We have reanalysed this question from the point of view of the
phenomenological time-dependent Ginzburg-Landau models. It was shown
that there are situations when one can go around the argument in \cite{hawu}.
A classic example are vortices in the Ginzburg-Landau models of the
fractional quantum Hall effect \cite{wl,glhe}. Chern-Simons interaction makes
them well localised objects and the dilute limit can be defined.
The main effect of the CS gauge field is to remove the divergent
gradient energies present in the global model. These divegencies are
also removed in the Ginzburg-Landau model for superconductors.
We have considered a gauged nonlinear Schrodinger equation as a minimal
version of time-dependent Ginzburg-Landau model. To model the structure
of the vortex core we assumed it was filled with normal fluid
so as to make the whole structure locally charge-neutral. The strenght of the
statistical interaction was proportional to the net deficit of the superfluid
replaced inside the core by the normal phase.

  It is argued \cite{baps} that the time evolution of the condensate
is much faster then that of the normal fluid. Thus in the static
case the core is filled with the normal fluid but when the vortex
moves fast enough the normal fluid can not adjust itself and does
not follow the vortex motion. It is a crude approximation \cite{stoof}
and it does not invalidate our results. The statistical interaction
shows up in the adiabatic approximation which is quite an
oposite limit. In this limit we can assume the nonuniformity of the
normal fluid does follow the vortex motion. The effective Lagrangian
for vortex motion can be arranged term by term according to the powers
of vortex velocity. The statistical interaction together with the term
responsible for the Magnus force \cite{at} are the first terms in this
effective Lagrangian. The distortion of the normal fluid distribution from
the static one, which is at least of the first order in velocities, can
contribute but only to the term quadratic in velocities which is
the next to leading term. Thus at least for slow motion as compared
to characteristic velocities, when expansion in powers of velocities is
justified, the Magnus force and statistical interaction are not altered.

  One can rewrite the local BCS model in
terms of the gap-function field \cite{abts,stoof,aatz} but the effective
theory appears to be nonlocal - it contains derivatives of arbitrary order.
In this way one goes from description in terms of electronic
degrees of freedom to description in terms of Cooper pairs.
The latter is well suited for the bulk of the superconductor.
However in the core of the vortex we can expect some decoherence,
which phenomenologically might be decribed by normal fluid.
Whenever degrees of freedom other than those of Cooper pairs
come into play the description in terms of Cooper pairs may
happen to be irrelevant. One way of dealing with the problem
is to truncate the effective theory on the lowest order in derivatives
and introduce more or less explicitly something like a normal
component. This approach is limited by the poor knowledge
about the nature of the normal fluid. Another approach is to take
as many orders in derivatives in the effective theory as possible to describe
also the normal fluid in terms of the gap function. The problem is that
in practice one would have to cut the expansion at certain order
and there is no warranty that such a cut theory would be self-consistent.

   To go around these problems we will derive the statistical
interaction in the framework of the Bogolubov-de Gennes formalism
for pure samples at zero temperature. We will work in the quasi
two-dimensional regime of long parallel vortices. In the case of
superconducting layers thicker than $100\AA$ the penetration lenght
$\Lambda$ is still very close to the penetration lenght $\lambda$ in the
bulk superconductor. In this regime we expect modifications  to be rather
quantitative in nature then qualitative.
An important first step was done by Gaitan \cite{gaitan} in his derivation of
the Berry phase responsible for the Magnus force. In this paper we are going
to reanalize his derivation and then to extend the method to the case of two
well separated vortices. In the microscopic theory the language
of the normal and superfluid is not very fruitful. The distinction of the
states below the Fermi surface into localised bound states and scattering
states appears to be more natural. The distinction does not
influence the value of the Magnus force but it is crucial
for the statistical interaction. The scattering states are common to
all vortices while bound states can be identified with particular ones.

\section{ Preliminaries on vortex solution in BCS theory }

   The problem of the vortex solution in the BCS theory can be
conveniently posed within the Bogolubov-de Gennes formalism \cite{dg}. It
has not been completely solved although some qualitative fictures of the
solution are known \cite{bardeen}. We will restrict here to listing the basic
ingredients of the formalism.

   The Bogolubov equation, defined in the Nambu spinor space \cite{nambu},
is
\begin{equation}\label{2.10}
(E_{n}-\hat{H}_{BOG})
\left(\begin{array}{c}
 u_{n} \\
 v_{n}
\end{array}\right)=0 \;\;,
\end{equation}
where $\hat{H}_{BOG}$ is the Bogolubov hamiltonian
\begin{equation}\label{2.20}
\hat{H}_{BOG}=
\left[\begin{array}{cc}
-\frac{1}{2}(\mbox{\boldmath $\nabla$}-ie\mbox{\boldmath $A$})^{2}-E_{F} &
\Delta(\mbox{\boldmath $x$}) \\
\Delta^{\star}(\mbox{\boldmath $x$}) &
\frac{1}{2}(\mbox{\boldmath $\nabla$}+ie\mbox{\boldmath $A$})^{2}+E_{F}
\end{array} \right]  \;\;.
\end{equation}
$E_{F}$ is the Fermi energy.
$\Delta(\mbox{\boldmath $x$})=\Delta_{0}(r)\exp(-i\theta)$ denotes
the gap function of vortex solution. $\Delta_{0}(r)$ interpolates
between $0$ at the origin and a constant, which we will call
$\sqrt{\rho_{0}}$, at infinity. There are both positive and negative energy
solutions. If $(u_{n},v_{n})$ is an eigenstate with energy $E_{n}>0$,
then $(-v^{\star}_{n},u^{\star}_{n})$ is a solution with energy
$-E_{n}<0$. The equations (\ref{2.10}) have to be supplemented
by a self-consistency condition
\begin{equation}\label{2.25}
\Delta(\mbox{\boldmath $x$})=
g\sum_{n}u_{n}(\mbox{\boldmath $x$})v_{n}^{\star}(\mbox{\boldmath $x$}) \;\;,
\end{equation}
where $g$ is the BCS coupling constant, together with Maxwell equations
determining the vector potential.

  The field operator for Nambu quasiparticles can be expanded in terms
of the solutions of Eq.(\ref{2.10}).
\begin{equation}\label{2.30}
\Psi( \mbox{\boldmath $x$} )=\left(\begin{array}{c}
                      \psi_{\uparrow}(\mbox{\boldmath $x$})   \\
                      \psi_{\downarrow}^{\dagger}(\mbox{\boldmath $x$})
                      \end{array}\right)=
\sum_{n}
\left[\begin{array}{c}
   \gamma_{n\uparrow}
   \left(\begin{array}{c} u_{n}(\mbox{\boldmath $x$}) \\
                          v_{n}(\mbox{\boldmath $x$}) \end{array}\right) +
   \gamma_{n\downarrow}^{\dagger}
   \left(\begin{array}{c} -v^{\star}_{n}(\mbox{\boldmath $x$}) \\
                           u^{\star}_{n}(\mbox{\boldmath $x$})
   \end{array}\right)
\end{array}\right] \;\;.
\end{equation}
The negative energy states are occupied in the BCS ground state
\begin{equation}\label{2.40}
\mid BCS > = \prod_{n} \gamma_{n\downarrow} \mid 0 > \;\;.
\end{equation}
The eigenstates satisfy the following orthogonality
\begin{eqnarray}\label{2.50}
&&\int d^{3}x\;
  [u_{n}(\mbox{\boldmath $x$})u^{\star}_{m}(\mbox{\boldmath $x$})
  +v_{n}(\mbox{\boldmath $x$})v^{\star}_{m}(\mbox{\boldmath $x$})]=
                                                   \delta_{nm}
\;\;,\nonumber\\ &&\int d^{3}x\;
  [u_{n}(\mbox{\boldmath $x$})v_{m}(\mbox{\boldmath $x$})
  -v_{n}(\mbox{\boldmath $x$})u_{m}(\mbox{\boldmath $x$})]=0
\end{eqnarray}
and completeness relations
\begin{eqnarray}\label{2.60}
&&\sum_{n}[u_{n}(\mbox{\boldmath $x$})
                               u_{n}^{\star}(\mbox{\boldmath $x$}^{\prime})
          +v_{n}^{\star}(\mbox{\boldmath $x$})
                               v_{n}(\mbox{\boldmath $x$}^{\prime})]
  =\delta(\mbox{\boldmath $x$}-\mbox{\boldmath $x$}^{\prime})
\;\;,\nonumber\\ &&\sum_{n}[u_{n}(\mbox{\boldmath $x$})
                       v_{n}^{\star}(\mbox{\boldmath $x$}^{\prime})
          -v_{n}^{\star}(\mbox{\boldmath $x$})
                               v_{n}(\mbox{\boldmath $x$}^{\prime})]=0 \;\;.
\end{eqnarray}
With the help of these relations the creation and annihilation operators
can be expressed as
\begin{eqnarray}\label{2.70}
&&\gamma_{n\downarrow}=\int d^{3}x\;
[-\psi^{\dagger}_{\uparrow}(\mbox{\boldmath $x$})
                                v_{n}^{\star}(\mbox{\boldmath $x$})
 +\psi_{\downarrow}(\mbox{\boldmath $x$})
                                u^{\star}_{n}(\mbox{\boldmath $x$})] \;\;,
                                                                 \nonumber\\
&&\gamma_{n\uparrow}^{\dagger}=\int d^{3}x\;
[\psi^{\dagger}_{\uparrow}(\mbox{\boldmath $x$})
                                u_{n}^{\star}(\mbox{\boldmath $x$})
  +\psi_{\downarrow}(\mbox{\boldmath $x$})
                                v_{n}(\mbox{\boldmath $x$})] \;\;.
\end{eqnarray}
Adiabatic vortex motion \cite{gaitan} gives rise to a Berry phase
in the solutions of Eq.(\ref{2.10}),
$(u_{n},v_{n})\rightarrow \exp[i\phi_{n}](u_{n},v_{n})$.
These Berry phases sum up to the total Berry phase picked up
by the ground state $\mid BCS >\rightarrow \exp[i\Gamma] \mid BCS >$
which with the help of Eqs.(\ref{2.70},\ref{2.40}) can be established
to be
\begin{equation}\label{2.80}
\Gamma=-\sum_{n}\phi_{n} \;\;.
\end{equation}
To persue some of the questions we need more detailed knowledge
about the eigenstates. The axially symmetric ansatz takes the form
\begin{equation}\label{2.90}
\chi_{n}(\mbox{\boldmath $x$})=\left ( \begin{array}{c}
                              u_{n}(\mbox{\boldmath $x$}) \\
                              v_{n}(\mbox{\boldmath $x$})
                              \end{array} \right )=
e^{ik_{z}z} e^{i(\mu-\frac{1}{2}\sigma_{z})\theta} f_{n}(r) \;\;,
\end{equation}
where $\hbar k_{z}$ is the $z$-component of the momentum. $\mu$ must
be a half-integer for the expression to be single-valued. The functions
$f_{n}(r)$ have been investigated in \cite{bardeen} with the help of WKB
approximation. We will quote some more detailed results in the following.

\section{ Origin of the Magnus force }

   Now we are going to rederive the Berry phase responsible for
the Magnus force following the argument of Gaitan \cite{gaitan}. In
comparison with \cite{gaitan} we clarify some points and remove some
unnecessary assumptions.

   The general form of the Berry phase in two dimensions is
\begin{equation}\label{3.10}
\phi_{n}=i\int dt\; \int d^{2}x\;
\chi_{n}^{\dagger} (\frac{d}{dt}+i\frac{e}{\hbar}A_{0}) \chi_{n} \;\;.
\end{equation}
The time derivative is understood as a total derivative with respect
to slow degrees of freedom. For an adiabatic motion of a
single vortex the derivative has to be replaced by
$\dot{\mbox{\boldmath $r$}}_{0}\mbox{\boldmath $\nabla$}
_{\mbox{\boldmath $r_{0}$}}$, where $\mbox{\boldmath $ r_{0} $}$
is a position of the vortex singularity. The scalar potential
vanishes for the vortex solution so we will skip the second term in
what follows. With the axially symmetric ansatz (\ref{2.90}) the phase
becomes
\begin{equation}\label{3.20}
\phi_{n}=\int dt\;\int d^{2}x\; \dot{ \mbox{\boldmath $r$} }_{0} [
   (\chi^{\dagger}_{n}(-\mu+\frac{1}{2}\sigma_{z})\chi_{n})
   \mbox{\boldmath $ \nabla $}_{\mbox{\boldmath $r_{0}$}}  \theta+
   f^{\dagger}_{n}(r) \mbox{\boldmath $ \nabla $}_{\mbox{\boldmath $r_{0}$}}
                                                                   f_{n}(r)]
\end{equation}
The second term vanishes by symmetry arguments. The contribution of the first
term to the total Berry phase is
\begin{equation}\label{3.30}
\Gamma=-\sum_{n}\phi_{n}=-\int dt\;\int d^{2}x\;
    ( \dot{ \mbox{\boldmath $r$} }_{0} \mbox{\boldmath $ \nabla $}
    _{ \mbox{\boldmath $ r_{0} $} }\theta) S(r)
\end{equation}
where
\begin{equation}\label{3.40}
S(r)=\sum_{n}[\mid u_{n}(r)\mid^{2}(-\mu+\frac{1}{2})+
              \mid v_{n}(r)\mid^{2}(-\mu-\frac{1}{2})] \;\;.
\end{equation}
$\hbar S$ is minus the z-component of the canonical angular momentum density.
It is not a gauge-invariant integral of motion. $S$ is the expectation
value density of the operator $-i\hbar\frac{\partial}{\partial\theta}$
instead of
the gauge-invariant
$-i\hbar\frac{\partial}{\partial\theta}+\sigma_{z}A_{\theta}$.

$S(0)=0$ because either $u_{n}$ ($v_{n}$)
or the factor $(-\mu\stackrel{+}{-}\frac{1}{2})$ vanishes at the origin.
To find out its asymptotic behavior at infinity we would need a much more
detailed knowledge about the solutions. We go around this problem by
resorting to the effective theory which is
equivalent to the microscopic formalism. The general term linear in
the covariant time derivative reads
\begin{equation}\label{3.50}
  \int dt\;\int d^{2}x\;
  i[\Delta^{\star}(\hbar\partial_{t}+2ieA_{0})\Delta-c.c.]
                  [G(\Delta^{\star}\Delta)+spatial\;derivative\;terms]\;\;.
\end{equation}
By "$spatial\;derivative\;terms$" we mean terms which are of at least
first order in the covariant spatial derivatives. $G$ is a function
of $\Delta^{\star}\Delta$ only which tends to $\rho_{s}/\rho_{0}$ as the gap
function $\Delta$ approaches its asymptotic equilibrium value. $\rho_{s}$
is the equilibrium Cooper pairs' density.
Let us consider the adiabatic rotation of the vortex solution
$\Delta=\Delta_{0}(r)e^{-i\theta}$ around its axis,
$\theta\rightarrow\theta-\omega t$. The action picks up a term
(to lowest order in $\omega$)
\begin{equation}\label{3.60}
 \omega \int dt\;\int d^{2}x\; \{-\hbar\Delta^{\star}\Delta
       [G(\Delta^{\star}\Delta)+spatial\;derivative\;terms]\} \;\;.
\end{equation}
The spatial integral is just the total angular momentum.
For large $r$, where $A_{0}$ tends to zero, the density of this angular
momentum is, by gauge invariance (\ref{3.50}), equal to $\hbar$ times
minus the bulk Cooper pairs' density
$\hbar S\approx-\hbar\rho_{s}=\hbar\lim_{r\rightarrow\infty}
\frac{\delta W}{\delta (2eA_{0})}$, where $W$ is the effective action
and $\rho_{s}>0$. We have to stress that we make use of the effective
theory only very far from the vortex core where it should be equivalent
to the microscopic treatement. In particular in the distant asymptotic
region there is no contribution from unpaired bound states which
can not be described in terms of Cooper pairs.

   We have all we need to calculate the Berry phase. Let us expand
the integrand in Eq.(\ref{3.30}) around the vortex position
$\mbox{\boldmath $r_{0}$}=(X,Y)$ close to the origin, $(X,Y)=(0,0)$,
\begin{eqnarray}\label{3.70}
&&S=S(r)-S^{\prime}(r)[X\cos\theta+Y\sin\theta]
                                              +O(r_{0}^{2}) \;\;,\nonumber\\
&&\dot{ \mbox{\boldmath $r$} }_{0}\mbox{\boldmath $\nabla$}
  _{\mbox{\boldmath $r_{0}$}}\theta=
  \dot{X}(\frac{\sin\theta}{r}+\frac{X\sin 2\theta-Y\cos 2\theta}{r^{2}})+
  \dot{Y}(-\frac{\cos\theta}{r}-\frac{X\cos 2\theta+Y\sin 2\theta}{r^{2}})+
                                               O(r_{0}^{2}) \;\;.
\end{eqnarray}
A straightforward integration yields
\begin{equation}\label{3.80}
 \Gamma=-\pi [S(\infty)-S(0)] \int dt\;\varepsilon_{kl}X^{k}\dot{X}^{l}
                                                          +O(r_{0}^{2}) \;\;.
\end{equation}
The expression $O(r_{0}^{2})$ does vanish. We are considering single vortex
in absence of any driven current. Such a system is translationally
invariant and isotropic. The first term on the R.H.S. of Eq.(\ref{3.80})
is already the most general term linear in velocity which is, up to a total
time derivative, translationally invariant and isotropic. Thus we do not need
to consider finite $\mbox{\boldmath $r_{0}$}$ to obtain a generally valid
expression. The phase (\ref{3.80}) is remarkably simple to evaluate. For
a vortex with winding number $-1$ it reads
\begin{equation}\label{3.90}
 \Gamma=\pi\rho_{s}\int dt\; \varepsilon_{kl}X^{k}\dot{X}^{l} \;\;.
\end{equation}
 From our derivation of the Magnus force it is clear that
the Wess-Zumino term (in the gauge $A_{0}=0$)
\begin{equation}\label{3.100}
\int dt\;\int d^{2}x [\rho_{s}\partial_{t}\theta] \;\;,
\end{equation}
with $\rho_{s}=const$, does not make much sense as it stands.
$\rho_{s}$ is the same at the origin as at infinity so the Magnus force
vanishes. The formula (\ref{3.100})
is to be understood with an implicit assumption that a small area around
the phase singularity is excluded from the spatial integration. In
other words $\rho_{s}$ must be put equal to $0$ in this area.
It is not difficult to realise, by performing radial integration first
and then integration over the angle around the singularity, that the
way of regularisation does not matter. In particular it does not need
to be rotationally symmetric. The only factors that determine the Magnus
force are the two limit values of $\rho_{s}$. Thus vortices in a
condensate will always feel the Magnus force. It is not the case
for say Jackiw-Pi solitons \cite{jp}, where $\rho_{s}$ is zero both at
the origin and at infinity.

\section{ Mutual statistical interaction of vortices }

   Let us consider two vortices: "1" at the origin and "2"
very far apart at $\mbox{\boldmath $R$}(t)$. It is important to realise
that the eigenstates of the Bogolubov hamiltonian can be divided
into common scattering states, which we will still denote by just
$u_{n},v_{n}$, and bound states which can be identified with a given vortex
$u^{(1,2)}_{n},v^{(1,2)}_{n}$. Vortices are very distant so there is no
overlap between their localised bound states.

The bound states of the stationary vortex "1" feel what
is going on around them through the pair potential
$\Delta(t,\mbox{\boldmath $x$})$ inside and around the core.
Vortices are well localised so a fairly good
approximation to a two-vortex gap function is the product ansatz
\begin{equation}\label{4.10}
\sqrt{\rho_{0}}\Delta(t,\mbox{\boldmath $x$})
                =\Delta_{v}( \mbox{\boldmath $x$} )
                 \Delta_{v}[ \mbox{\boldmath $x-R$}(t) ] \;\;,
\end{equation}
where $\Delta_{v}(\mbox{\boldmath $x$})=\Delta_{0}(r)\exp(-i\theta)$ denotes
the gap function of a single vortex centered at the origin. Close to
$\mbox{\boldmath $x$}=0$ this expression can be further simplified
\begin{equation}\label{4.20}
\Delta(t,\mbox{\boldmath $x$})=\Delta_{v}(\mbox{\boldmath $x$})
                      e^{ -i\theta[\mbox{\boldmath $x-R$}(t)] } \;\;,
\end{equation}
Thus the bound states of the static vortex have to be modified as
\begin{equation}\label{4.30}
\chi_{n}^{(1)}[\mbox{\boldmath $x,R$}(t)] =
  e^{-i\frac{\sigma_{z}}{2}\theta[\mbox{\boldmath $x-R$}(t)]}
                             \chi_{n}(\mbox{\boldmath $x$}) \;\;.
\end{equation}
Their contribution to the Berry phase is
\begin{eqnarray}\label{4.40}
-\int dt\;\int d^{2}x\; \{ \dot{ \mbox{\boldmath $R$} }
       \mbox{\boldmath $\nabla$}
        _{\mbox{\boldmath $R$}}\theta[\mbox{\boldmath $x-R(t)$}] \}
 \sum_{bound\;st.}(\frac{1}{2}\mid u_{n}\mid^{2}-
                     \frac{1}{2}\mid v_{n}\mid^{2})\approx  \nonumber\\
 \{\frac{1}{2}\int d^{2}x\;
           \sum_{bound\;st.}(\mid u_{n}\mid^{2}-\mid v_{n}\mid^{2})\}
 \int dt\; \dot{ \mbox{\boldmath $R$} }\mbox{\boldmath $\nabla$}
             _{\mbox{\boldmath $R$}}\theta[\mbox{\boldmath $R(t)$}]\;\;,
\end{eqnarray}
where the approximate equality is valid for small
$\frac{r_{c}}{R}$, where $r_{c}$ is a radius of the core.
The equality $\mid u_{n}\mid^{2}=\mid v_{n}\mid^{2}$ holds for the bound
states, at least up to the WKB approximation \cite{bardeen},
so their contribution to the Berry phase vanishes.

  Now as the vortex "2" moves its bound states follow its trajectory
$\chi^{(2)}[\mbox{\boldmath $r-R$}(t)]$, similarly as in the single vortex
case
considered in the previous section. In addition, as an effect due to the
vortex "1" (\ref{4.10}), their components perform the relative phase rotation
\begin{equation}\label{4.50}
\chi_{n}^{(2)}[\mbox{\boldmath $x,R$}(t)] \approx
  e^{-i\frac{\sigma_{z}}{2}\theta[\mbox{\boldmath $R$}(t)]}
  \chi_{n}[\mbox{\boldmath $x-R(t)$}] \;\;.
\end{equation}
The contribution from this relative phase rotation is once again zero.
The bound states contribute but only to the Magnus term (\ref{3.90}) just
as in the single vortex case.

   The bound states do not give rise to any new effects so let us consider
scattering states common to both vortices. In the vortex core region
the asymptotes of the scattering states must be close to those
of the scattering states for a single vortex but the phase has to be
replaced by the asymptote of the phase in the product ansatz (\ref{4.20}).
Close to the origin
\begin{equation}\label{4.60}
 \chi_{n}[\mbox{\boldmath $x,R$}(t)]\approx e^{ik_{z}z}
   e^{i(\mu-\frac{\sigma_{z}}{2})
   \{\theta(\mbox{\boldmath $x$})+\theta[\mbox{\boldmath $x-R$}(t)]\}}
   f_{n}(r) \;\;.
\end{equation}
The contribution from around the stationary vortex is
\begin{equation}\label{4.70}
-\int dt\;\int d^{2}x\;
 \dot{\mbox{\boldmath $R$}}\mbox{\boldmath $\nabla$}_{\mbox{\boldmath $R$}}
  \theta[\mbox{\boldmath $x-R$}(t)]  \bar{S}(r) \;\;,
\end{equation}
where $\bar{S}$ is a part of the canonical angular momentum due to
the scattering states
\begin{equation}\label{4.80}
\bar{S}(r)=\sum_{scatt.\;st.}[\mid u_{n}(r)\mid^{2}(-\mu+\frac{1}{2})+
              \mid v_{n}(r)\mid^{2}(-\mu-\frac{1}{2})] \;\;.
\end{equation}
Far from the core $\bar{S}\approx S\approx -\rho_{s}$.
If $\bar{S}$ were equal to $-\rho_{s}$ also in the core, the contribution to
the Berry phase from (\ref{4.70}) would be just the same as to the
Magnus term. Thus we are interested only in the effects due
to deviations of $\bar{S}$ from its asymptotic value $-\rho_{s}$.
Inside the core $\mid u_{n}\mid^{2}$ and $\mid v_{n}\mid^{2}$
are changed by a factor which is $>1$ for the states with $\mu$ negative
and $<1$ for $\mu$ positive \cite{bardeen}. The net deviation
$\delta\bar{S}(r)=\bar{S}(r)+\rho_{s}$ is positive.
At the very origin $\delta\bar{S}(0)=\rho_{s}$. The total change in the Berry
phase is twice that in (\ref{4.70}), as there are two vortices, and
amounts to
\begin{equation}\label{4.90}
\delta\Gamma=[2\int d^{2}x\;\delta\bar{S}(r)]
                \int dt\; \dot{\mbox{\boldmath $R$}}\mbox{\boldmath $\nabla$}
                _{\mbox{\boldmath $R$}}\theta(\mbox{\boldmath $R$})\;\;.
\end{equation}
Thus the total Berry phase for a dilute vortex system is
\begin{equation}\label{4.100}
\Gamma=\int dt\;
     [ -\pi\rho_{s}\sum_{p} n_{p}\varepsilon^{kl}X^{k}_{(p)}\dot{X}^{l}_{(p)}+
       \alpha\sum_{p<q}n_{p}n_{q}\frac{d}{dt}\Theta_{(p,q)}]  \;\;,
\end{equation}
where the indices $p,q$ run over vortices, $n$'s are their winding numbers,
$\Theta_{(p,q)}$ is the angle between the $p$-th and $q$-th vortex
and the numerical factor $\alpha$ can be read from Eq.(\ref{4.90})
\begin{equation}\label{4.110}
\alpha=2 \int d^{2}x\; \delta\bar{S}(r) \;>\; 0  \;\;.
\end{equation}
$\alpha$ is roughly the number of electrons inside the core
and as such it can range from $\sim 1$ for high $T_{c}$ superconductors
to $\sim 10^{5}$ for some conventional type $II$ superconductors.

\section{ Vortex statistics within variational wave-function approach }

   Once we have derived statistical interaction in the microscopic
setting it may be worthwhile to reanalise some earlier approaches
to similar problems. In the paper by Ao and Thouless \cite{at} the Magnus
force was derived with the help of the variational many-electron
vortex wave-function
\begin{equation}\label{at.10}
\psi_{v}[z]=\exp[\frac{i}{2}\sum_{k}\theta(z_{k}-z_{0})]\psi_{0}[z] \;\;,
\end{equation}
where $z_{0}$ is a complex vortex position, $z_{k}$'s are positions
of electrons and $\psi_{0}[z]$ is an antisymmetric variational function.
The phase factors in the wave-function are determined by the demand
of correct electronic quantumstatistics and by topological properties.
There are variational profile functions in $\psi_{0}$ which can not be
established without dynamical considerations. One can consider an adiabatic
vortex motion along some trajectory and calculate the Berry phase
picked up by the wave-function. This Berry phase coinsides with
Eq.(\ref{3.90}). For a closed path the Berry phase is proportional
to the number of electrons enclosed by the trajectory.

This setting is convenient to analise what is the dependence of
the Magnus force on impurities \cite{at}. An impurity can be viewed as an
attractive potential which traps some of electrons in localised
bound states. The trapped electrons disappear from the ansatz (\ref{at.10}).
The Berry phase is still proportional to the area enclosed by the trajectory
but this time the area should not be multiplied by the total density
of electrons but rather by the total density minus the density of electrons
trapped by impurities. Impurities lower the value of the Magnus force.

Now let us consider the effect of an exchange of two vortices.
More precisely, let us fix the position of one vortex and consider
another distant vortex moving around it. One could
argue there is no special effect because the net charge of any vortex
must be zero. Provided the trajectory is large enough, there is no
change in the number of enclosed electrons due to the enclosed vortex.
The last sentence is certainly true but the example with impurities
tought us that it is not the total number of electrons that really
matters but rather the number of electrons in the coherent state
described by the wave-function (\ref{at.10}). We know from the discussion
in the previous sections that inside vortex core the scattering
or continuum states are replaced by bound states.
Thus vortex can be viewed as a kind of impurity, which traps some
of electrons into localised bound states with energies within the energy gap
band. The localised electrons are removed from the wave-function
(\ref{at.10}). There is an additional Berry phase proportional
to the number of vortices enclosed by the trajectory. Each enclosed
vortex contributes a term proportional to the number of electrons
trapped inside its core.

   We can consider a path for a chosen vortex in a more or less uniform
distribution of vortices. If we neglect possible intervortex correlation
effects, the background vortices could be regarded as uniform distribution
of impurities lowering the density of electrons in the coherent state
(\ref{at.10}). In this mean-field approximation the Magnus force
acting on a choosen vortex is lowered by the presence of another vortices.
This approximation is nothing else but the delocalisation procedure
so often applied to anyonic systems. The Magnus force can be interpreted
as Lorenz force due to interaction of effectively charged vortices with some
uniform effective magnetic field. The statistical interaction can be seen as
Aharonov-Bohm effect due to the fluxes attached to vortices. In the
mean-field approximation the fluxes, which are opposite to the external
flux, are delocalised and they lower the net uniform flux. In the same
way the real impurities can be interpreted as localised fluxes, opposite
to the external field, randomly distributed over the plane. If the M-F
approximation appears to work for real impurities, it will also work
for vortices.

\section{ Hall angle and vortex density }

  The fact that the value of the Magnus force can be lowered with
increasing density of vortices can, in principle, be observable
in Hall experiments \cite{hl}. The vortex density should increase
and the M-F Magnus force should decrease with increasing real magnetic field.
This should manifest itself in the changes of the measured Hall
angle. Vortex equation of motion takes the form \cite{ah}
\begin{equation}\label{at.20}
m_{eff}\ddot{\mbox{\boldmath $r$}}=
\frac{\rho_{s}hd}{2}
( \dot{\mbox{\boldmath $r$}} - \mbox{\boldmath $v_{s}$} )\times
  \hat{\mbox{\boldmath $z$}}
-\eta d \dot{\mbox{\boldmath $r$}} + \mbox{\boldmath $F_{pin}$}
+ \mbox{\boldmath $f$} \;\;,
\end{equation}
where $m_{eff}$ is a small effective vortex mass, $\eta$ is a vortex
viscosity, $\mbox{\boldmath $F_{pin}$}$ is a pinning force,
$\mbox{\boldmath $f$}$ is a fluctuating force, $d$ is a sample thickness
and $\mbox{\boldmath $v_{s}$}$ is a driven uniform superfluid velocity.
When we neglect pinning and average over fluctuations the stationary state
motion will be determined by the equation
\begin{equation}\label{at.30}
( \dot{\mbox{\boldmath $r$}} - \mbox{\boldmath $v_{s}$} )\times
  \hat{\mbox{\boldmath $z$}}
=\tan(\theta_{H})\dot{\mbox{\boldmath $r$}} \;\;,
\end{equation}
where $\tan(\theta_{H})=\frac{2\eta}{\rho_{s}h}$. The solution is
\begin{equation}\label{at.40}
\dot{\mbox{\boldmath $r$}}=
\frac{ \mbox{\boldmath $v_{s}$}
    + (\hat{\mbox{\boldmath $z$}}\times\mbox{\boldmath $v_{s}$})
    \tan(\theta_{H}) }{1+\tan^{2}(\theta_{H}) }  \;\;.
\end{equation}
$\theta_{H}$ is the angle between the superfluid velocity
$\mbox{\boldmath $v_{s}$}$ and the stationary vortex velocity.
The angle is the larger the weaker is the Magnus force.
If the effective Magnus force is lowered with increased vortex density
the angle should also grow with external magnetic field, which
drives the rise in vortex density. The changes of the Magnus force
due to changes in vortex density should be the more rapid the larger is
the number of electrons trapped in the vortex core.
For this reason we would recommend experiments on mildly type $II$
conventional superconductors with large vortex cores
(large correlation length $\xi$). Rather strong viscosity
should be prefered for
the angle to be more sensitive to the strengh of the Magnus force.
The sample should be pure of pinning centers to avoid obscure pinning
effects.

\section{ Vortices' fractional quantum Hall effect  }

  To summarise our knowledge about the dynamics of planar vortices,
let us write down an effective Lagrangian for diluted vortices with
topological charge $-1$
\begin{equation}\label{5.10}
L_{eff}=\sum_{p} [\frac{1}{2}m_{eff}\dot{\mbox{\boldmath $X$}}_{p}
                                    \dot{\mbox{\boldmath $X$}}_{p}
       +\hbar\pi\rho_{s} \mbox{\boldmath $X_{p}$}
        \times \dot{ \mbox{\boldmath $X$} }_{p}  ]
       +\sum_{p<q}[\hbar\alpha\frac{d}{dt}\Theta_{(p,q)}
       -V_{eff}(\mid\mbox{\boldmath $X_{p}-X_{q}$}\mid)] \;\;.
\end{equation}
The indices $p,q$ run over vortices. $m_{eff}$ is the effective
vortex mass. It is usually estimated to be around
$10^{8}m_{e}/m=10^{-2}m_{e}/\AA$.

  The second term in Eq.(\ref{5.10}) decribes interaction of vortices
with effective magnetic field. We stress that this field has nothing to do
with the real
magnetic field $B_{ext}$, which in this case is just a device
to drive the changes of vortex density. If we assumed the density
of electrons to be $\sim 10^{30} m^{-3}=1\AA^{-3}$ the effective magnetic
field defined by $\frac{eB_{eff}}{2}=\pi\hbar\rho_{s}$ would turn out to
be $\sim 10^{6} T/\AA$. When compared with the effective vortex mass
per $1\AA$ the magnetic field turns out to be incredibly strong.
Its effect on a vortex should be the same as that of the $10^{8}T$ magnetic
field on an electron. Vortices can be expected to be confined to the
lowest Landau level (LLL).

  Now let us consider a single vortex at $z=X_{1}+iX_{2}$. What is the
magnetic lenght $l$ which determines the size of the LLL wavepacket
\begin{equation}\label{5.20}
\psi_{0}(z,\bar{z})=
            \frac{1}{\sqrt{2\pi l^{2}}}e^{-\frac{z\bar{z}}{4l^{2}} }\;\;?
\end{equation}
The magnetic lenght is determined by the strength of the exactly known
Magnus force $l^{2}=\frac{1}{2\pi\rho_{s}}$. The area over which
the center of such a vortex fluctuates can be estimated to be
$\frac{2}{\rho_{s}}$, which is of the same order as the area per
one electron. This effect is significant for extremely type $II$
superconductors where the core is very thin.

  Now we are prepared to address the question of the fractional Hall effect.
Vortices are anyons with a statistical parameter $\alpha$. A nontrivial
statistics would be sufficient to prevent them from overlapping if
the third term in (\ref{5.10}) were not regularised at short distances
and replaced by mutual charge-flux interaction \cite{kld}. Fortunately we
also have an effective short range mutual repulsion $V_{eff}$ which at low
temperatures may be sufficient to keep vortices at a distance.
On the other hand the potential is very weak as compared to the Landau
energy so the mixing with higher LL's should be in any case negligible.
Following Laughlin \cite{laughlin}, the trial wave-function for a many
vortex state can be written
\begin{equation}\label{5.30}
\psi_{m}[z]=\prod_{p<q}(z_{p}-z_{q})^{\alpha+2m}
           \prod_{r} exp(-\mid z_{r} \mid^{2}/4l^{2}) \;\;,
\end{equation}
where $z_{p}$'s are positions of vortices. $m$ is a nonnegative integer
so that the exponent $(\alpha+2m)$ provides a correct quantumstatistics.
The density $n$ and the filling factor $\nu$ in such a state are
\begin{equation}\label{5.40}
\nu_{m}=2\pi l^{2} n_{m}=\frac{1}{\alpha+2m} \;\;.
\end{equation}
For $m=0$ the density is just $n_{0}=1/2\pi\alpha l^{2}=\rho_{s}/\alpha$.
$\alpha$ is roughly equal to the number of electrons
inside the core so in the $m=0$ state vortex cores
would have to overlap slightly. Certainly this density is close
to that of the Abrikosov lattice \cite{abr}. For larger $m$ the density
is smaller and finally we should get outside of the crystalline
regime. Then as the density of vortices is driven to change by the changes
of the real magnetic field $B_{ext}$ we should observe some plateaux
at the densities $n_{m}$ and maybe also at some other quantised
filling factors. For conventional superconductors, which are mildly type $II$,
$\alpha$ is large, so the difference between $n_{m}$ and $n_{m+1}$
is too small for the plateaux to be distinguishable
experimentally. For extremaly type $II$ superconductors or high $T_{c}$
superconductors we can expect $\alpha$ to be even $\sim 1$.
In such a case the first plateaux can be expected already
at $\sim 1/3$ and $\sim 1/5$ of the Abrikosov lattice density.

   The main observation of this section is that if the Magnus force
is translated into the language of interaction with some effective uniform
magnetic field,
the field appears to be even $\sim 10^{6}T/\AA$. It has nothing
to do with the real magnetic field which is at best $\sim 10 T$.
The FQHE is a result of this huge effective
magnetic field and of repulsive intervortex potential but its existence
does not depend on the vortex statistics as the FQHE is possible even for
bosons \cite{jain}. However the pattern of the FQHE plateaux can help to
identify the quantumstatistics.

   An experimental setup to detect the FQHE would consist of a planar sample
of some pure superconducting material in external uniform perpendicular
magnetic field. One would have to measure the total
flux $\Phi$, which penetrates through the sample. This flux gives the
actual number of vortices in the sample because each vortex carries one
flux quantum.
Just after continuously turning on weak magnetic field
the field lines would be pushed out of the sample. Only above
some threshold value of the flux the energy of the field could be
minimised by creating a vortex line. The story would repeat until
some stable FQHE plateau were reached. The plateau is a manifestation of
a very stable many-vortex state so an addition of one vortex to this state
would cost more energy then to a state far from the plateau. When the stable
state is approached from low densities the addition of one more vortex
should be much easier than usually. Quite opposite, when we lower
the external magnetic field it is easier to remove one flux quantum
from the sample just above the stable state and more difficult to remove
it just below the state, see the figure. The moduli of the change in the
driving flux $\Delta\Phi$ (or some equivalent quantity) necessary to change
the flux through the sample by one quantum will develop a characteristic
pattern around the stable density $n_{0}$. Two measurements, one with
adiabatically increasing and one with adiabatically
decreasing external magnetic field,
would give two curves with opposite polarisation. Taking their difference
will amplify the effects due to the plateau and remove the not neccesarily
constant bias.

  It should be stressed here that our results are not in contradiction
with predictions of a bosonic Hall effect in Josephson junction arrays
\cite{jja}. The arrays are strictly planar devices. The penetration
lenght $\Lambda$ is likely to be greater then the sample size.
What is more, it seems to be  possible to excite a vortex without any bound
states in the core, which indeed might be a boson. Our results are
exact in the limit of long straight-linear parallel vortices. In practice
it is sufficient that the sample is thick enough for the penetration
lenght $\Lambda$ to be close to that in the bulk. The sample thickness
would have to be a bit more then $d=100\AA$. For such a $d$ there is
still no space for vortex entanglement. In the absence of entanglement
vortices can be uniquely projected on a plane. In nonzero temperature
one should expect transversal modes to be excited. These excitations
do not affect the Berry phase picked up during vortices exchange as it
depends on the overall topological properties of vortices. Vortices
excited to different transversal states although in principle distinguishable
interact statistically. It is an example of mutual statistical interaction
\cite{mut} introduced first to describe interlayer phase correlations
in the double-layer Hall effect. The plateaux pattern for our anyonic
vortices is given by (\ref{5.40}). In distinction to bosonic Hall
effect the constant $\alpha$ is nonzero.

\section{ Summary }

We have given two new arguments why vortices in superconducting
films should be anyons. In addition we have discussed two
different experiments where this theoretical prediction could
be verified.

\paragraph{Aknowledgement.}

I would like to thank Prof. B.Halperin for stimulating comments on
Ref.\cite{anyon}. This research was supported in part by the KBN grant
No. 2 P03B 085 08 and in part by Foundation for Polish Science fellowship.


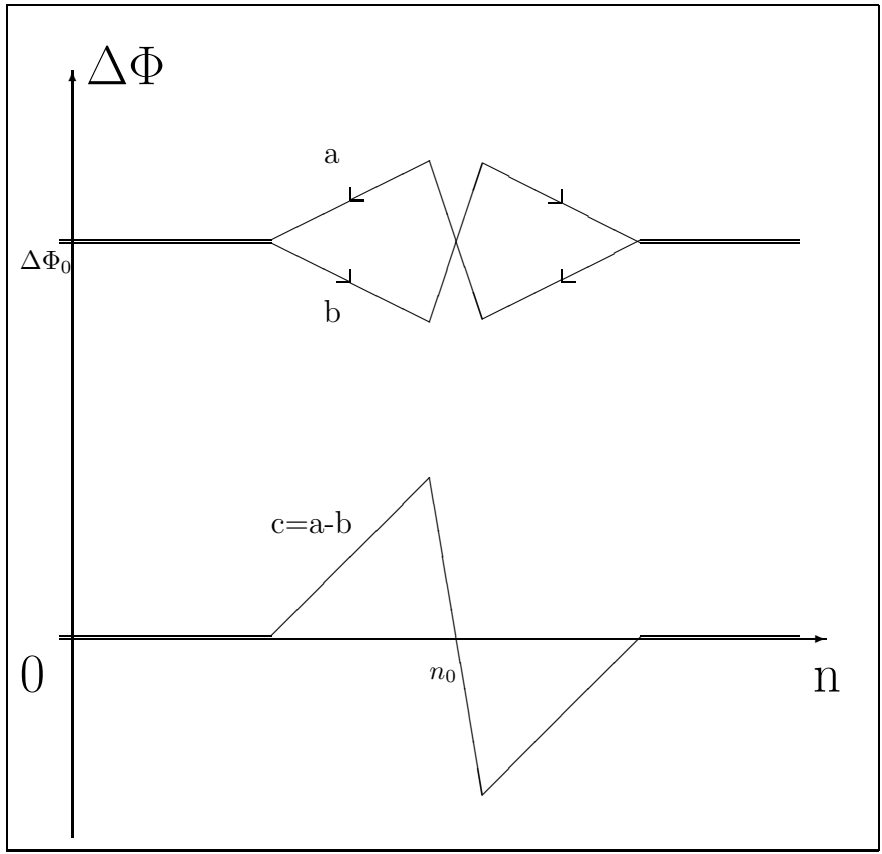
\paragraph{FIGURE CAPTION.}
Detection of the Hall plateau. $\Delta\Phi$ is the moduli of the change in
the external magnetic flux necessary to drive the change in the number of
vortices by one.
$\Delta\Phi_{0}$ is its average value far from the plateau. $n$ is the
density of vortices and $n_{0}$ its value at the plateau. The curve "a"
coresponds to decreasing while "b" to increasing number of vortices. The
curve "c" is the difference between "a" and "b".

\begin{picture}(340,330)(25,0)
\put(5,80){\vector(1,0){290}}
\put(10,5){\vector(0,1){290}}
\put(5,81){\line(1,0){80}}
\put(85,81){\line(1,1){60}}
\put(145,141){\line(1,-6){20}}
\put(165,21){\line(1,1){60}}
\put(225,81){\line(1,0){60}}
\put(5,230){\line(1,0){80}}
\put(85,230){\line(2,-1){60}}
\put(145,200){\line(1,3){20}}
\put(165,260){\line(2,-1){60}}
\put(225,230){\line(1,0){60}}
\put(5,231){\line(1,0){80}}
\put(85,231){\line(2,1){60}}
\put(145,261){\line(1,-3){20}}
\put(165,201){\line(2,1){60}}
\put(225,231){\line(1,0){60}}
\put(115,215){\line(0,1){5}}
\put(115,215){\line(-1,0){5}}
\put(115,246){\line(0,1){5}}
\put(115,246){\line(1,0){5}}
\put(195,245){\line(-1,0){5}}
\put(195,245){\line(0,1){5}}
\put(195,215){\line(0,1){5}}
\put(195,215){\line(1,0){5}}
\put(290,60){\mbox{\huge n}}
\put(15,290){\mbox{\huge $\Delta\Phi$}}
\put(-10,60){\mbox{\huge 0}}
\put(-10,220){$\Delta\Phi_{0}$}
\put(105,200){\mbox{\large b}}
\put(105,260){\mbox{\large a}}
\put(85,120){\mbox{\large c=a-b}}
\put(145,65){$n_{0}$}
\put(-15,0){\line(1,0){330}}
\put(315,0){\line(0,1){320}}
\put(315,320){\line(-1,0){330}}
\put(-15,0){\line(0,1){320}}
\end{picture}

\end{document}